\documentclass[aps,prb,twocolumn,superscriptaddress,showpacs,floatfix]{revtex4}
\usepackage{ulem}
\usepackage{epstopdf}
\usepackage{graphicx}
\usepackage{dcolumn}
\usepackage{bm}
\usepackage{amsfonts}
\usepackage{amsmath}
\usepackage{color}

\newcommand{\etal}{\textit{et al.}}

\begin{document}
\baselineskip=0.5cm
\renewcommand{\thefigure}{\arabic{figure}}
\def\be{\begin{equation}}
\def\ee{\end{equation}}
\def\ber{\begin{eqnarray}}
\def\eer{\end{eqnarray}}

\def\sigmav{\mbox{\boldmath $\sigma$}}
\def\tauv{\mbox{\boldmath $\tau$}}

\title{Anisotropic RKKY interaction in spin polarized graphene}
\date{\today}

\author{F. Parhizgar}
\affiliation{School of Physics, Institute for Research in
Fundamental Sciences (IPM), Tehran 19395-5531, Iran}
\author{Reza Asgari}
\email{asgari@ipm.ir} \affiliation{School of Physics, Institute
for Research in Fundamental Sciences (IPM), Tehran 19395-5531,
Iran}
\author{Saeed H. Abedinpour}
\affiliation{Department of Physics, Institute for Advanced Studies
in Basic Sciences (IASBS), Zanjan 45137-66731, Iran }
\author{M. Zareyan}
\affiliation{Department of Physics, Institute for Advanced Studies
in Basic Sciences (IASBS), Zanjan 45137-66731, Iran }

\begin{abstract}
We study the Ruderman-Kittle-Kasuya-Yosida (RKKY) interaction in
the presence of spin polarized two dimensional Dirac fermions. We
show that a spin polarization along the $z$-axis mediates an
anisotropic interaction which corresponds to a XXZ model
interaction between two magnetic moments. For undoped graphene,
while the $x$ part of interaction keeps its constant ferromagnetic
sign, its $z$ part oscillates with the distance of magnetic
impurities, $R$. A finite doping causes that both parts of the
interaction oscillate with $R$. We explore a beating pattern of
oscillations of the RKKY interaction along armchair and zigzag
lattice directions, which occurs for some certain values of the
chemical potential. The two characteristic periods of the beating
are determined by inverse of the difference and the sum of the
chemical potential and the spin polarization.
\end{abstract}

\pacs{75.30.Hx, 75.10.Lp,75.10.Jm, 75.30.Ds} \maketitle

\section{Introduction}

The charge and spin oscillatory interactions in metals has
attracted considerable attention both on the theoretical and
experimental sides~\cite{vignale, wu}. Ruderman and
Kittle~\cite{ref:rk} suggested that the spin oscillatory
interaction in metals could provide a long-range interaction
between nuclear spins in metals. Afterwards, Kasuya and
Yosida~\cite{ref:ky} extended the theory to include the long-range
interaction between magnetic impurities and thus the combined
refers to RKKY interaction.

The recent discovery of graphene~\cite{novoselov}, the
two-dimensional crystal of carbon atoms, has provided a new
material with a peculiar structure for the charge and the spin
interactions. This stable crystal has already attracted
considerable attention because of its unusual effective many-body
properties~\cite{bostwick_science_2010, yafis, polini, polini2,ee
interaction, asgari_im, kotov_arXiv_2010, qaiumzadeh} that follow
from chiral nature of linearly dispersing low energy excitations
described by pair of Dirac cones at the $K$ and $K'$ edges of the
first Brillouin zone.

The RKKY interaction in pristine graphene has been studied by
several groups~\cite{ref:all,ref:saremi, ref:sherafati}. Due to
the particle-hole symmetry of graphene, the RKKY interaction
induces ferromagnetic correlation between magnetic impurities on
the same sublattice, while anti-ferromagnetic correlation between
the ones on different sublattices. The dependence of the
interaction on the distance $R$ between two local magnetic
moments, at the Dirac point, is found to be $R^{-3}$, whereas it
behaves as $R^{-2}$ in conventional two-dimensional (2D)
systems~\cite{bergmann}. Such a fast decay rate denotes that the
interaction is rather short-ranged. In doped graphene, on the
other hand, the spacial dependence of the interaction is predicted
to be similar to conventional 2D systems, but this still remains
to be experimentally verified.

Due to the fact that the RKKY interaction is originated by the
exchange coupling between the impurity moments and the spin of
itinerant electrons in the bulk of the system, spin polarization
of electrons is expected to influence directly this interaction
\cite{Zarand_PRL_2002}. In particular, combination of the
spin-dependence with a Dirac-like spectrum can mediate a much
richer collective behavior of magnetic
adatoms~\cite{Liu_PRL_2009}. This has been explained for surface
states of a three dimensional topological insulator, on which
magnetic impurities exhibit a frustrated RKKY interaction with two
possible phases of ordered ferromagnetic phase and a disordered
spin glass phase~\cite{Abanin_PRL_2011}. Graphene, in particular,
with imbalanced chemical potentials of spin-up and spin-down
electrons, presents a unique spin chiral material in which the
interplay between the spin polarization, gapless spectrum, and the
chiral nature of electrons have been shown to result intriguing
phenomena~\cite{MZ_Others_2008-2011}. In two-dimension graphene
system, the polarization of the chemical potential can be tuned to
be of order or even higher than the mean chemical potential, the
condition which is not possible in ordinary conductors.  The aim
of the present study is to address the question of how this
peculiarity can affect the collective coupling of magnetic
impurities on the surface of graphene sheet with a finite
polarization of the spin.

In this work, we calculate the RKKY interaction mediated by
spin-polarized Dirac fermions in a monolayer graphene using the
Green's function method. Our theory for the spin polarization
dependence of RKKY interaction is motivated not only by
fundamental transport considerations, but also by application and
potential future experiments in graphene spintrotic field. With a
spin-polarization along the $z$-axis, we show that the RKKY
interaction is anisotropic corresponding to a XXZ model
interaction between the two magnetic moments when their spin
orientations are fixed. Besides $R^{-3}$ dependence of the
interaction for undoped graphene, we show particularly that the
interaction behaves like $R^{-2}$ when the spin polarization is
finite. In addition, a beating pattern for the interaction in the
cases where impurities are located along certain directions is
obtained near the resonance condition which is controlled by the
chemical potential and the spin polarization.

The paper is organized as the following. In Sec.~\ref{sect:theory}
we introduce the formalism that will be used in calculating the
RKKY interaction from the lattice Green's function. In
Sec.~\ref{sect:results} we present our analytic and numeric
results for the coupling strengths of the RKKY interaction in both
undoped and doped graphene sheets. Sec.~\ref{sect:conclusion}
contains discussions and a brief summary of our main results.

\section{Method and Theory}\label{sect:theory}

 We consider a spin-polarized graphene system identified by a spin dependent
chemical potential $\mu_s$ ($s=\pm1$), implying a mean chemical
potential $\mu=\sum_{s} \mu_s/2$ and the spin polarization
$\mu_p=\sum_{s}s \mu_s/2$. Such a spin-polarization can be
injected, for instance, by ferromagnetic electrodes on top of the
graphene sheet
~\cite{Dugaev_Others_2006-2008,Tombros_Nature_2007}. Intrinsic
ferromagnetic correlations are also predicted to exist in graphene
sheets~\cite{Peres_PRB_2005} and nanoribbons with zigzag edges
~\cite{Son_Nature_2006} under certain conditions.

The electronic structure of spin polarized graphene can be
reasonably well described using a rather simple tight-binding
Hamiltonian, leading to analytical solutions for their energy
dispersion and related eigenstates. The noninteracting nearest
neighbor tight-binding Hamiltonian for $\pi$-band electrons with
spin $s$, is determined by~\cite{gapsub1}
\begin{eqnarray}\label{eq:TB}
\mathcal{\hat{H}}^s_0=&-&t\sum_{\langle i,j\rangle}\left(a_{i,s}^\dagger b_{j,s}+b_{i,s}^\dagger a_{j,s}\right)\nonumber\\
&-&s\mu_p\sum_{i}\left(a_{i,s}^\dagger a_{i,s}+b_{i,s}^\dagger b_{i,s}\right)~,
\end{eqnarray}
where $a_{i,s}(b_{i,s})$ annihilates an electron with spin $s$ on
sublattice A(B) of unit cell $i$ and $t\simeq 2.9$ eV denotes the
nearest neighbor hopping parameter~\cite{ref:hopping}. The sum
$\langle i,j \rangle$ in Eq.~(\ref{eq:TB}) runs over distinct
nearest neighbors.

The $s$-component of the noninteracting Hamiltonian in momentum
space is written as
\begin{eqnarray}
\mathcal{\hat{H}}^s_0=\begin{pmatrix}
-s\mu_p & f({\bf k}) \\ f^*({\bf k}) & -s\mu_p
\end{pmatrix}~,
\end{eqnarray}
where the form factor in general case is $f({\bf
k})=-t\sum_{j}e^{i{\bf k}\cdot {\bf d}_j}$, in which ${\bf d}_j$'s
are nearest neighbor position vectors. In this work, we are
interested in the low-energy behavior, in which $f({\bf k})=v_{\rm
F} k \Phi(k)$, where $\Phi(k)=e^{i(\pi/3+\theta_k)}$ at the Dirac
point $K$ and $\Phi(k)=-e^{i(\pi/3-\theta_k)}$ at the another
Dirac point, $K'$, the chiral angle is
$\theta_k=\tan^{-1}(k_x/k_y)$, $v_{\rm F}=3ta/2\hbar\simeq10^6$
m/s is the Fermi velocity with $a\simeq1.42${\AA} being the
Carbon-Carbon distance in honeycomb lattice.

Our system incorporates two localized magnetic moments whose
interaction is mediated through a spin polarized electron liquid.
We assume that the graphene is spin polarized first, and then we
add the magnetic moments. The contact interaction between the spin
of itinerant electrons and two magnetic impurities with magnetic
moments ${\bf M_1}$ and ${\bf M_2}$, located respectively at ${\bf
R}_1$ and ${\bf R}_2$, is given by
\begin{equation}
\mathcal{\hat{H}}_{\rm int} = \lambda \sum_{j=1,2} {\bf M}_j\cdot  {\bf s}({\bf R}_j)~,
\end{equation}
where $\lambda$ is the coupling constant between conduction
electrons and impurity, ${\bf s}({\bf r})=\frac{\hbar}{4} \sum_i
\delta ({\bf r}_i-{\bf r}) \sigmav_i$ is the spin density
operator~\cite{vignale} with ${\bf r}_i$ and $\sigmav_i$ being the
position and vector of spin operators of $i$th electron.

The RKKY interaction which arises from the quantum effects is
obtained by using a second order
perturbation~\cite{ref:rk,ref:ky,ref:Imamura} which reads as (from now on we set $\hbar=1$)

\begin{eqnarray} \label{eq:RKKY}
\mathcal{\hat {H}}^{\alpha\beta}_{\rm RKKY}&=&
\frac{-\lambda^2}{\pi}\Im m
\int_{-\infty}^{\infty}d\varepsilon
{\rm Tr}[({\bf M}_1\cdot \sigmav)G_{\alpha\beta}({\bf R}_1,{\bf R}_2;\varepsilon) \nonumber\\
&\times& ({\bf M}_2 \cdot \sigmav) G_{\beta\alpha}({\bf R}_2,{\bf R}_1;\varepsilon)] n(\mu)~,
\end{eqnarray}
where $n(\mu)$ denotes the Fermi-Dirac distribution function,
$\sigmav$ is the vector of Pauli matrices in the {\it spin} space,
$G_{\alpha\beta}({\bf R}_1,{\bf R}_2;\varepsilon)$ is a $2\times
2$ matrix of the single particle retarded Green's function in spin
space, and $\alpha$ and $\beta$ refer to the sublattices where two
impurities are placed and finally, the trace is taken over the
spin degree of freedom.

For spin unpolarized graphene, Eq.~(\ref{eq:RKKY}) simplifies to
$\mathcal{\hat {H}}^{\alpha\beta}_{\rm RKKY}=\frac{\lambda^2}{4}
\chi({\bf R}_1,{\bf R}_2) ~  {\bf M}_1 \cdot {\bf M}_2$, where
$\chi({\bf R}_1,{\bf R}_2)$ is the spin susceptibility of the
itinerant electrons and determines the indirect interaction
between two local moments.

In order to calculate the interaction Hamiltonian of
Eq.~(\ref{eq:RKKY}), the form of the electronic single particle
Green's function, $G^s({\bf R}_1,{\b R}_2;\varepsilon)=<{\bf
R}_1|(\varepsilon+i0^{+}-\mathcal{\hat {H}}^s_0)^{-1}|{\bf R}_2>$
is needed. To calculate the retarded Green's function in real
space, its Fourier components in momentum space might be first
obtained. Due to the fact that our 2D Dirac fermion system is
noninteracting and thus the direction of spin remains unchanged,
the retarded Green's function $G_{\alpha\beta}$ are diagonal in
the spin space \be\label{green1} G^s_{AA}({\bf R},0,\varepsilon)=(
e^{i{\bf K}\cdot {\bf R}}+e^{i{\bf K'}\cdot {\bf R}} )
g_{AA}(\varepsilon-s\mu_p)~, \ee and
\begin{eqnarray}\label{green2}
G^s_{AB}({\bf R},0,\varepsilon)=&e^{i\pi/3}(e^{i{\bf K}\cdot {\bf R}+i\theta_R}
-e^{i{\bf K'}\cdot {\bf R}-i\theta_R})\nonumber \\
&\times
g_{AB}(\varepsilon-s\mu_p)~.
\end{eqnarray}

Moreover, $G^s_{BB}=G^s_{AA}$ and $G^s_{BA}(0,{\bf
R},\varepsilon)=exp(-i\pi/3)(exp({-i{\bf K}\cdot {\bf
R}-i\theta_R}) -exp({-i{\bf K'}\cdot {\bf
R}+i\theta_R}))g_{AB}(\varepsilon-s\mu_p)$. Here
$g_{AA}(\varepsilon)=\gamma \varepsilon K_0(-i\varepsilon R/v_{\rm
F})$ and $g_{AB}(\varepsilon)=\gamma \varepsilon K_1(-i\varepsilon
R/v_{\rm F})$, where $K_0(x)$ and $K_1(x)$ are the modified Bessel
functions of the second kind, $\theta_R$ is the angle of the
position ${\bf R}$ with respect to the ${\bf K'}-{\bf K}$
direction and $\gamma=-2\pi/(\Omega v_{\rm F}^2)$, in which
$\Omega$ is the area of the Brillouin zone.

By inserting the retarded Green's functions given by
Eqs.~(\ref{green1}) and (\ref{green2}) in Eq.~(\ref{eq:RKKY}), and
taking the trace over spin degree of matrices, the RKKY
Hamiltonian simplifies to

\begin{equation}
\mathcal{H}^{\alpha\beta}_{\rm RKKY} =\frac{\lambda ^2}{\pi}
  [J^{\alpha\beta}_x (M_{1x}M_{2x}+M_{1y}M_{2y}) +J^{\alpha\beta}_{z} M_{1z}M_{2z}]~,
\end{equation}
which is the honored XXZ model. Here $J^{\alpha\beta}_x =-C
\Phi_{\alpha\beta}I^{\alpha\beta}_x/R^3$ and $J^{\alpha\beta}_z
=-C \Phi_{\alpha\beta} I^{\alpha\beta}_z/R^3$, with
$C=2(2\pi)^2/\Omega^2 v_{\rm F}$, $\Phi_{AA} = 1+\cos[({\bf
K}-{\bf K'})\cdot {\bf R}]$ and $\Phi_{AB}=-1+\cos[({\bf K}-{\bf
K'})\cdot{\bf R}+2\theta_R]$. The different components of
$I^{\alpha\beta}$ for impurities on the same sublattice read as
\be\label{eq:J}
\begin{split}
I^{AA}_x&=2\Im m\left[\int_{-\infty}^{y_{\rm F}}dy~x_{-}K_0(-ix_{-}) x_{+} K_0(-ix_{+})\right] ~,\\
I^{AA}_z&=\Im m\left[\int_{-\infty}^{y_{\rm F}}dy~\sum_{s=\pm} x_{s}^2 K_0^2(-ix_{s})\right]~,
\end{split}
\ee
where $y_{\rm F} = \mu R  /v_{\rm F}$, $x_{\pm}=y \pm h_{\rm F}$,
$y=\varepsilon R/ v_{\rm F}$, and $h_{\rm F} = \mu_p R /v_{\rm
F}$. For impurities on different sublattices, one only needs to
replace $K_0(x)$ with $ K_1(x)$ in the above equations. Note that
$J^{BB}=J^{AA}$ and $J^{BA}=J^{AB}$.

We find analytic results for the $z$ component of the RKKY
exchange coupling strength $I_z$, for both cases where magnetic
moments are located on the same or different sublattices. For the
same sublattice case, we begin by splitting the integral in the
second line of Eq.~(\ref{eq:J}) into the conduction and valance
band contributions, and find
\begin{eqnarray} \label{eq:Iz}
I^{AA}_z(R)=&2\Im m\left[\int_{-\infty}^0 dx\ x^2 K_0^2(-ix)\right]\nonumber\\
&+\sum_{s=\pm}\Im m\left[ \int_0^{x_{\rm Fs}} dx\ x^2 K_0^2(-ix)\right]~,
\end{eqnarray}
where $x_{F \pm }=y_{\rm F} \pm h_{\rm F}$. The first integral can
be solved~\cite{ref:sherafati,ref:saremi} easily and the result is
$\pi^2/32$. The contribution from the second line of
Eq.~(\ref{eq:Iz}), can be also obtained by replacing $\Im
m[K_0^2(-ix)]$ with $-\pi^2 sign(x)J_0(|x|)Y_0(|x|)/2$, and using
the following relation \be \int_0^{x_{\rm F}} dx\, x^2 sign(x)
J_0(|x|)Y_0(|x|)=-\frac{|x_{\rm F}|}{2\sqrt{\pi}} M(x_{\rm F})~,
\ee where
$M(x)=G[\{\{\frac{1}{2}\},\{\frac{3}{2}\}\},\{\{1,1\},\{-\frac{1}{2},1\}\},x^2]$
is Meijer $G$-function~\cite{ref:MeijerG}. As a result, the
function $I_z^{AA}$ is given by \be\label{eq:Iz_AA}
I_z^{AA}(R)=\frac{\pi^2}{4}\sum_{s=\pm}\left[\frac{1}{8}+\frac{|x_{{\rm
F}s}|}{\sqrt{\pi}} M(x_{{\rm F}s})\right]~. \ee

To calculate the long-range behavior of the RKKY interaction, the
asymptotic behavior of the Meijer $G$-function is needed. It is
also easy to see that asymptotic behavior of $M(x)$ at large $x$
is~\cite{ref:MeijerG}
$\left[2\cos(2x)+8x\sin(2x)-\pi\right]/(8\sqrt{\pi}x)$. It should
be noticed that the $M(x)$ tends to its long-range asymptotic
expression for $x > 2$. Therefore, $I_z^{AA}(R)$ for the long-range
regime is simplified as
\begin{equation}\label{eq:Iz_AA_lim}
I_z^{AA}(R\gg
a)\approx\frac{\pi}{16}\sum_{s=\pm}\left[\cos(2x_{Fs})+4x_{Fs}\sin(2x_{Fs})\right]
\end{equation}

On the other hand, for the case that the impurities are located on
two different sublattices we can follow the same procedure
discussed above, while we use $\int _{-\infty}^0 dx\,x^2 \Im
m[K_1^2(-ix)] = 3\pi^2/32$, and $\Im m[K_1^2(-ix)]=\pi^2
sign(x)J_1(|x|)Y_1(|x|)/2$ to find \be \int_0^{x_F} dx x^2 sign(x)
J_1(|x|) Y_1(|x|) =-\frac{|x_F|}{2\sqrt{\pi}} M'(x_F)~, \ee where
$M'(x)=G[\{\{\frac{1}{2}\},\{\frac{3}{2}\}\},\{\{1,2\},\{-\frac{1}{2},0\}\},x^2]$.
Finally the $I_z^{AB}(R)$ reads as \be\label{eq:Iz_AB}
I_z^{AB}(R)=\frac{\pi^2}{4}\sum_{s=\pm}\left[\frac{3}{8}-\frac{|x_{{\rm
F}s}|}{\sqrt{\pi}} M'(x_{{\rm F}s})\right]~. \ee The asymptotic
behavior of $M'(x)$ at large $x$ is
$[3\pi-10\cos(2x)-8x\sin(2x)]/(8\sqrt{\pi}|x|)$. Therefore, the
long-range behavior of $I_z^{AB}$ is obtained as
\be\label{eq:Iz_AB_lim} I_z^{AB}(R\gg
a)\approx\frac{\pi}{16}\sum_{s=\pm}\left[5\cos(2x_{{\rm
F}s})+4x_{{\rm F}s}\sin(2x_{{\rm F}s})\right]~.
 \ee

It should be mentioned that we were not able to find simple
analytic expressions for the in plane components of the exchange
coupling, $I^{\alpha\beta}_x$ and in the next section we will
present our numerical results of them.

\begin{figure}
\includegraphics[width=1.1\linewidth]{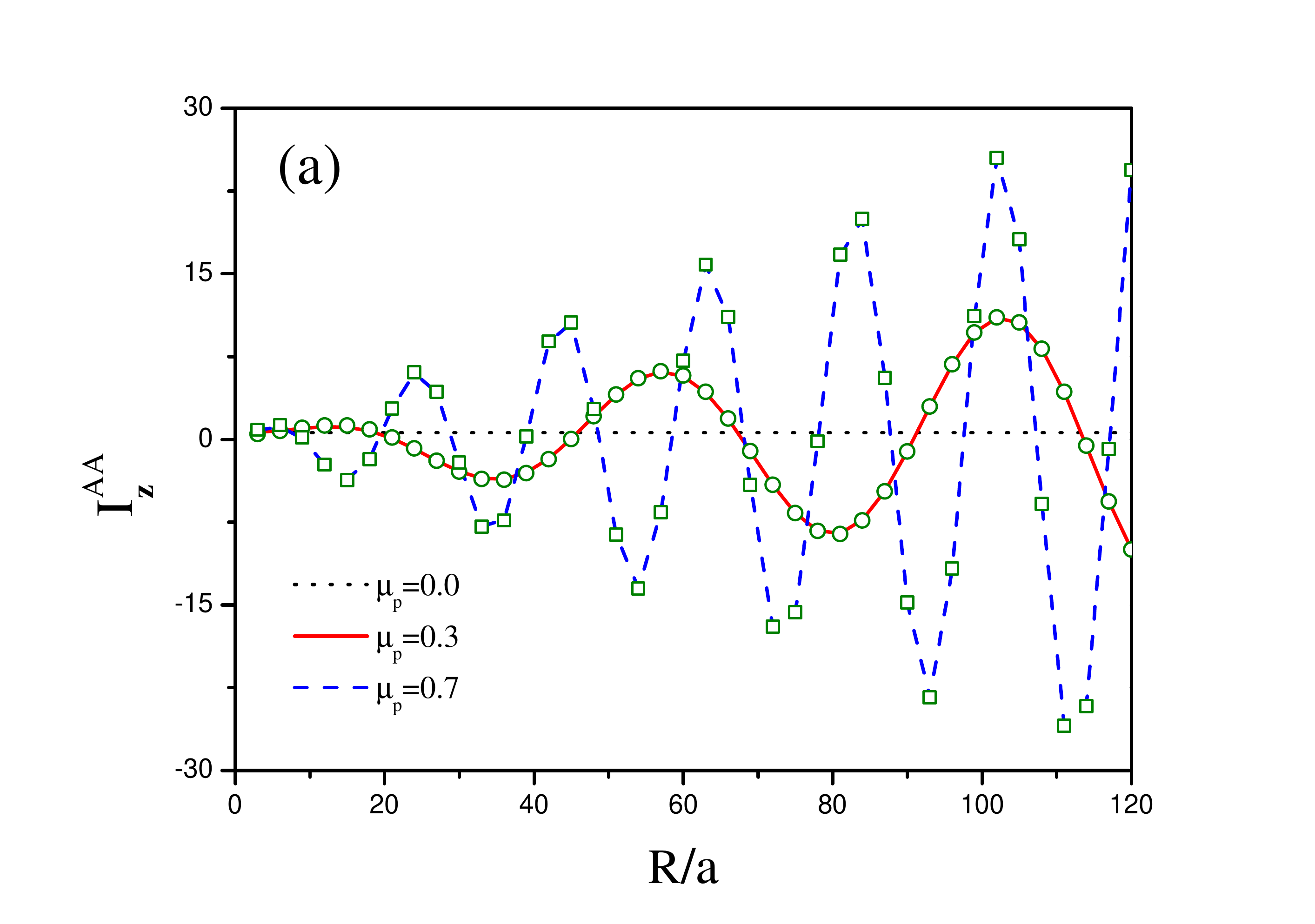}
\includegraphics[width=1.1\linewidth]{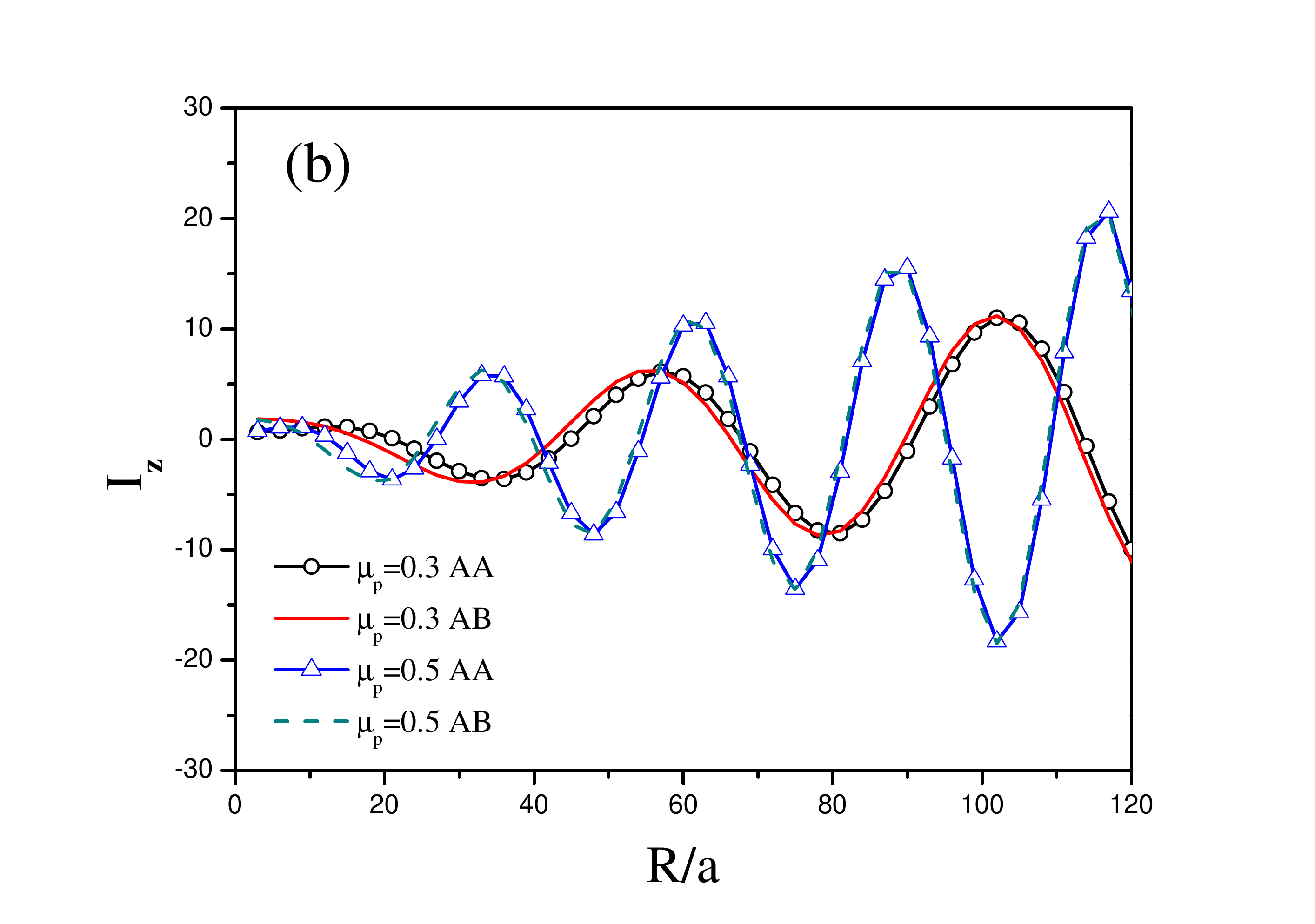}
\caption{(Color online) (a) The integral $I^{AA}_z$ as a function of the distance $R$ when both impurities are located on the same sublattice for various values of the spin polarization, $\mu_p$ in units of eV.
The chemical potential is set to zero. Symbols refer to the analytical results of Eq.~(\ref{eq:Iz_AA_lim})
which are compared to the numerical evaluation of Eq.~(\ref{eq:Iz_AA}), plotted as lines.
For $\mu_p=0$, $I^{AA}_z=I^{AB}_z$ is just a constant.
(b) A comparison between the integral $I^{AA}_z$ and $I^{AB}_z$ as a function of
the distance $R$ for various values of the spin polarization $\mu_p$ in units of eV.
At finite $\mu_p$, the integral $I_z$ has a quite
different behavior, oscillating as a function of $R$, with a period
given by $2\pi/h_{F}$ and a linearly growing amplitude. A
comparison between $R$-dependence of the integral $I^{AA}_z$ and
that of $I^{AB}_z$, shows their difference at short distance
while reaching each other as $R$ increases.
\label{fig:AAIzEF0}
}
\end{figure}

\section{Numerical Results and Discussions}\label{sect:results}

In this section, we present our main results for the RKKY exchange
coupling in the presence of a spin polarization Dirac fermions
along the $z$-axis by analyzing the above calculated integrals of
$I_x$ and $I_z$. We extend the previously studied
~\cite{ref:sherafati} results for dependencies on the distance $R$
and lattice direction $\theta_R$ to the case of $\mu_p\neq0$, for
two different regimes of undoped, ($\mu=0$) and doped, ($\mu \neq
0$) graphene.

\begin{figure}
\includegraphics[width=1.1\linewidth]{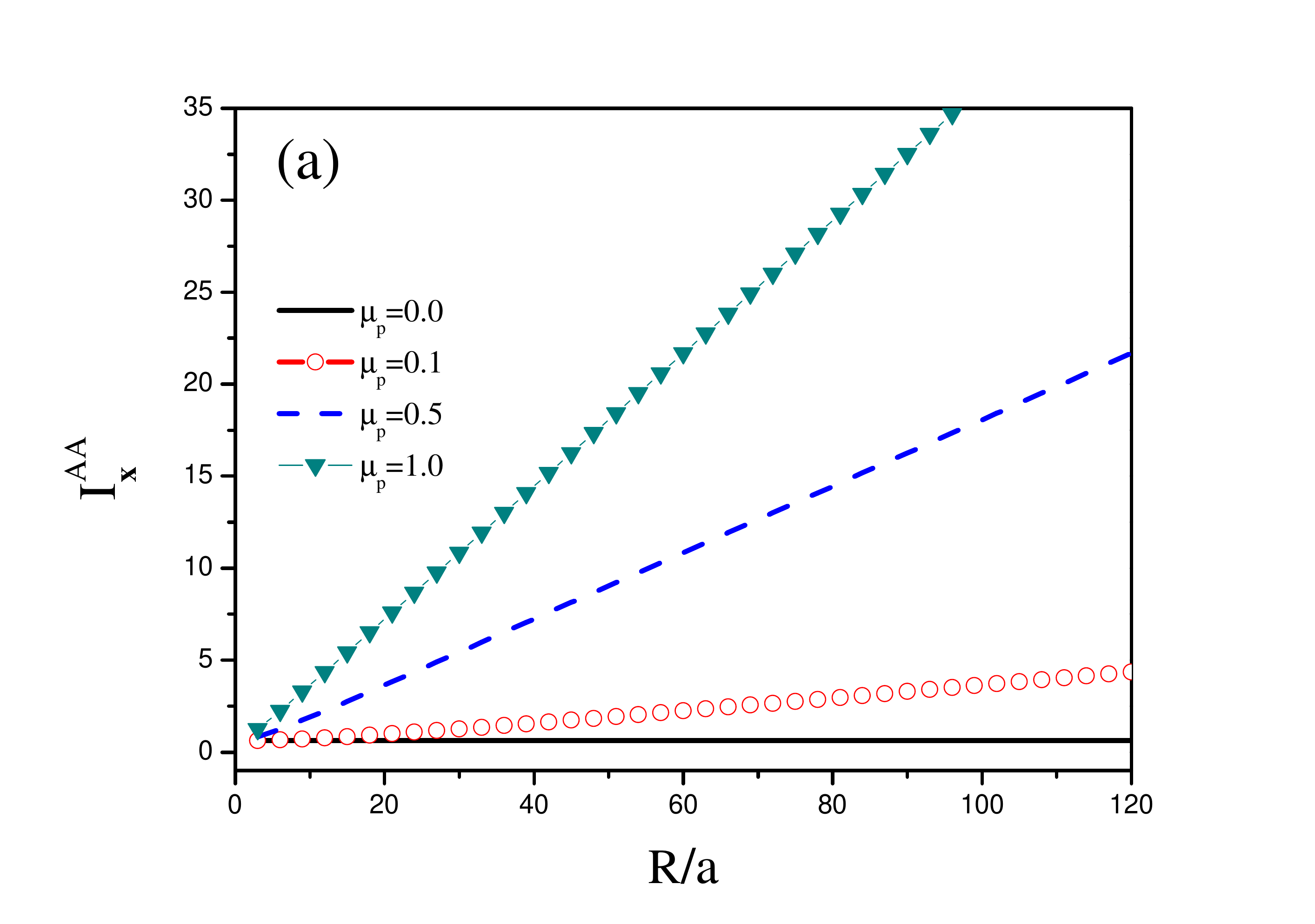}
\includegraphics[width=1.1\linewidth]{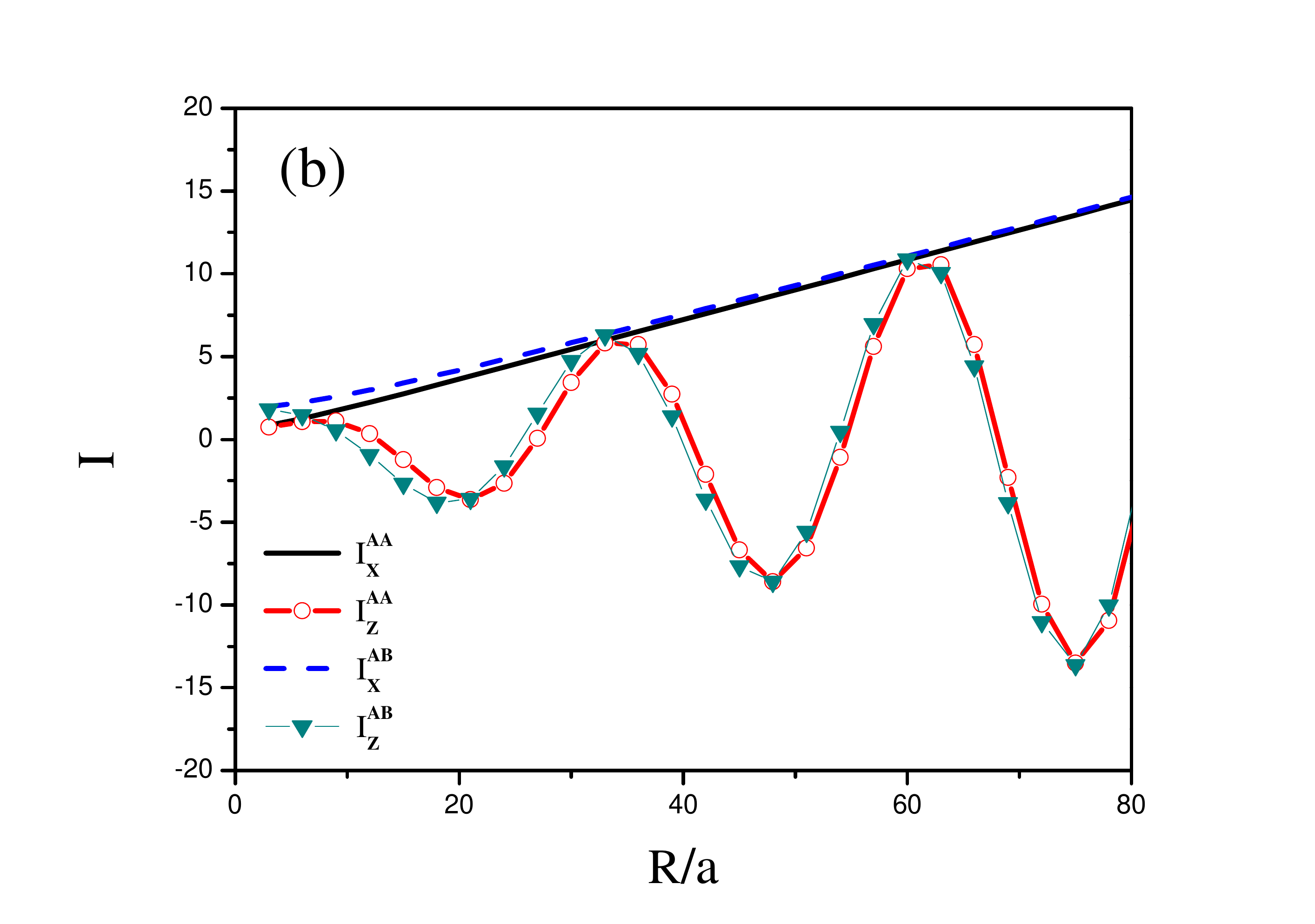}
\caption{(Color online) (a) The integral $I^{AA}_x$ versus distance between two impurities,
when both impurities are located on the same sublattice for undoped graphene and
for several values of $\mu_p$ in units of eV. Finite
$\mu_p$, produces a linear increase of $I^{AA}_x$ with
a slope proportional to $\mu_p$. (b) A comparison between
$I_x$ and $I_z$ for different configurations and for $\mu_p=0.5$ eV.
\label{fig:AAIxEF0} }
\end{figure}

For the $I_z$ component of interactions, we solve the two
expressions in Eqs.~(\ref{eq:Iz_AA}) and (\ref{eq:Iz_AB}),
numerically and then compare the results with asymptotic results
obtained from the analytical expressions given by
Eqs.~(\ref{eq:Iz_AA_lim} ) and (\ref{eq:Iz_AB_lim}), respectively.
Generally, the results obtained from the two approaches match
quite good in most of the case specially at long distances. The
distance dependence of $I_z$ for both $AA$ and $AB$ cases are
illustrated in Fig.~\ref{fig:AAIzEF0} for the undoped graphene.
For an unpolarized graphene $\mu_p=0$, $I^{AA}_z=I^{AA}_x$ is just
a constant. At finite $\mu_p$, the integral $I_z$ has quite
different behavior exhibiting an oscillatory behavior as a
function of $R$, with a linearly growing amplitude and a period
given by $2\pi/h_{F}$, as can be obtain directly from
Eq.~(\ref{eq:Iz_AA_lim}). This behavior of $I^{AA}_{z}$ results in
an oscillatory $J^{AA}_z$ with a decreasing amplitude as $R^{-2}$,
which mimics the behavior of the RKKY coupling of an unpolarized
doped graphene~\cite{ref:sherafati}. We can understand this
analogy by noting that the polarization induces spin-dependent
doping of up and down spin Dirac bands of an undoped sample by
shifting their chemical potential from the Dirac point. A
comparison between $R$-dependence of the integral $I^{AA}_z$ and
that of $I^{AB}_z$ for various values of $\mu_p$ in
Fig.~\ref{fig:AAIzEF0}(b), shows their difference at short
distance while reaching each other as $R$ increases. As
$R\rightarrow 0$ the coupling interactions tend to their values of
the unpolarized case where $J^{AB}$ is three times larger than
$J^{AB}$, as is discussed in Ref.~[\onlinecite{ref:sherafati}].

From the numerical calculations of the integrals appearing in
Eq.~(\ref{eq:J}), we can also obtain the behavior of $I^{AA}_x$
and $I^{AB}_x$ for the RKKY interaction coupling of the components
of the magnetic moments which are perpendicular to the spin
polarization axis. Fig.~\ref{fig:AAIxEF0}(a) shows $I^{AA}_x$ as a
function of $R$ for the undoped graphene at different values of
the spin polarization, $\mu_p$. For $\mu_p=0$, this function is a
constant resulting a $J^{AA}_x$ which decays as $R^{-3}$. A finite
difference $\mu_p$ between the chemical potentials of spin up and
spin down carriers, produces a linear increase of $I^{AA}_x$ with
a slope proportional to $\mu_p$. Thus, $J^{AA}_x$ decays as
$R^{-2}$. Importantly, the sign of interaction $J^{AA}_x$ is
always positive which shows that the coupling between the
perpendicular components of the moments remains ferromagnetic-like
for all $R$s. To analyze the difference between the two
configurations of $AA$ and $AB$, in Fig.~\ref{fig:AAIxEF0}(b) we
have compared $I^{AA}_x$ and $I^{AB}_x$, which shows that despite
the difference at short distances, they tend to each other at
larger distances.

\begin{figure}
\includegraphics[width=1.1\linewidth]{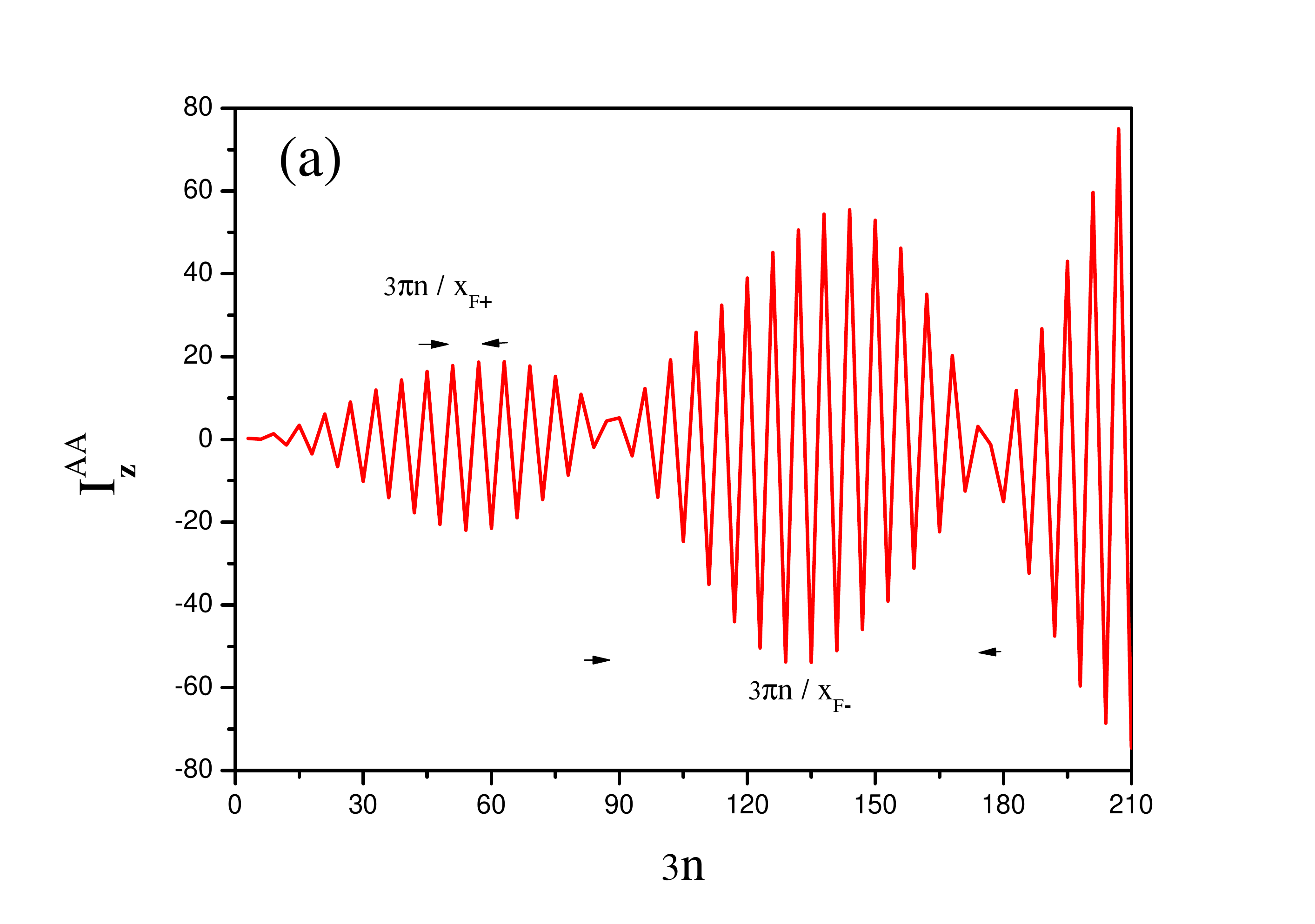}
\includegraphics[width=1.1\linewidth]{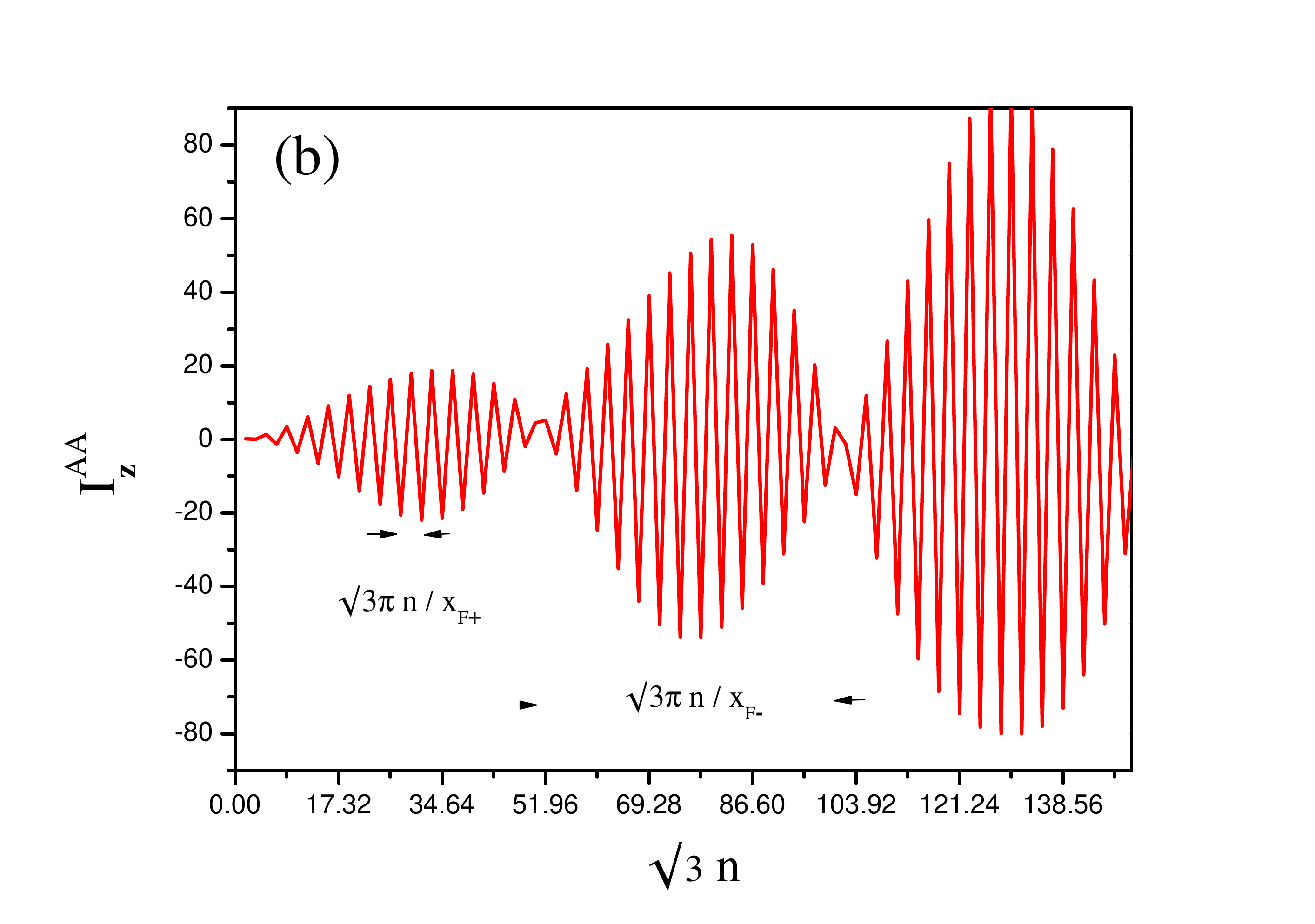}
\caption{(Color online) (a) The integral $I^{AA}_z$ as a function of the impurities distance
along armchair direction when
both impurities are located on the same sublattice. The existence of two
different periods in doped polarized graphene for certain values of
$\mu=1.2$ eV and $\mu_p=1.0$ eV is clear in this figure.
(b) The integral $I^{AA}_z$ as a distance
along zigzag direction for $\mu=1.2\sqrt{3}$ eV and $\mu_p=\sqrt{3}$ eV.}\label{fig:AAIzEF1}
\end{figure}

At finite values of both the chemical potential and the spin
polarization, more complicated behavior of the RKKY coupling can
be occurred. In this case, the behavior of $I_z$ is determined by
a superposition of four sinusoidal functions with two different
periods of $2\pi R/(x_{F-})$ and $2\pi R/(x_{F+})$ each occurring
twice with amplitudes $1, x_{F-}$ and $1,x_{F+}$, respectively. As
the result, we observe that for certain values of $\mu$ and
$\mu_p$, oscillations of $I_z$ exhibit a beating pattern with two
characteristic periods. Fig.~\ref{fig:AAIzEF1}(a) shows this
beating behavior of integral, $I^{AA}_z$ as a function of the
impurities distance along armchair direction (where  $R=3na$ with
$n$ being an integer number) for $\mu=1.2$ eV and $\mu_p=1.0$ eV.
Fig.~\ref{fig:AAIzEF1}(b) shows the similar behavior of integral
$I^{AA}_z$ as a distance along zigzag direction (where
$R=\sqrt{3}ma$ for an integer $m$) for $\mu=1.2\sqrt{3}$ eV and
$\mu_p=\sqrt{3}$ eV. We have obtained a similar beating pattern
for oscillations of $I^{AB}_z$, which also occurs for a certain
values of $\mu$ and $\mu_p$.

\begin{figure}
\includegraphics[width=1.1\linewidth]{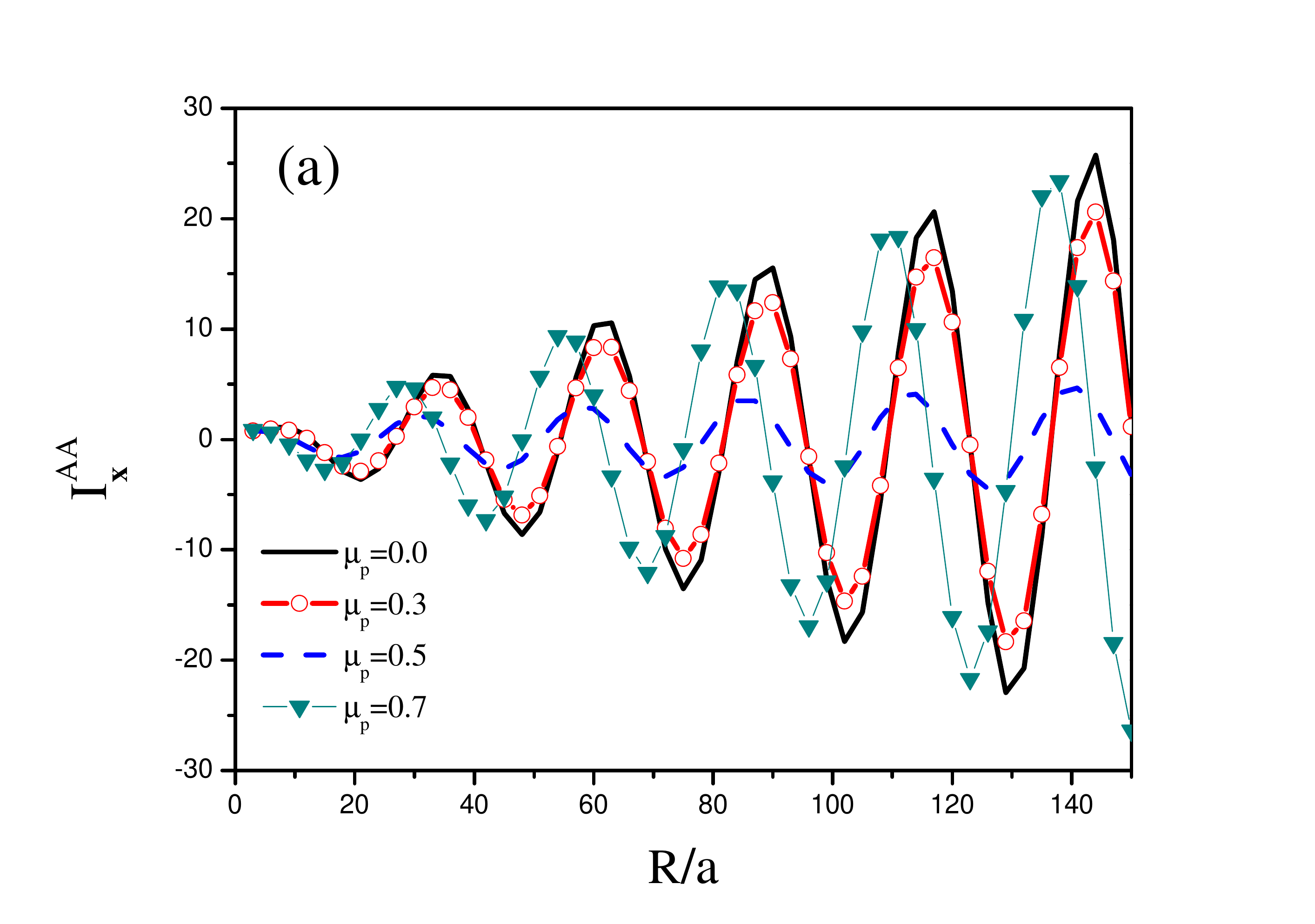}
\includegraphics[width=1.1\linewidth]{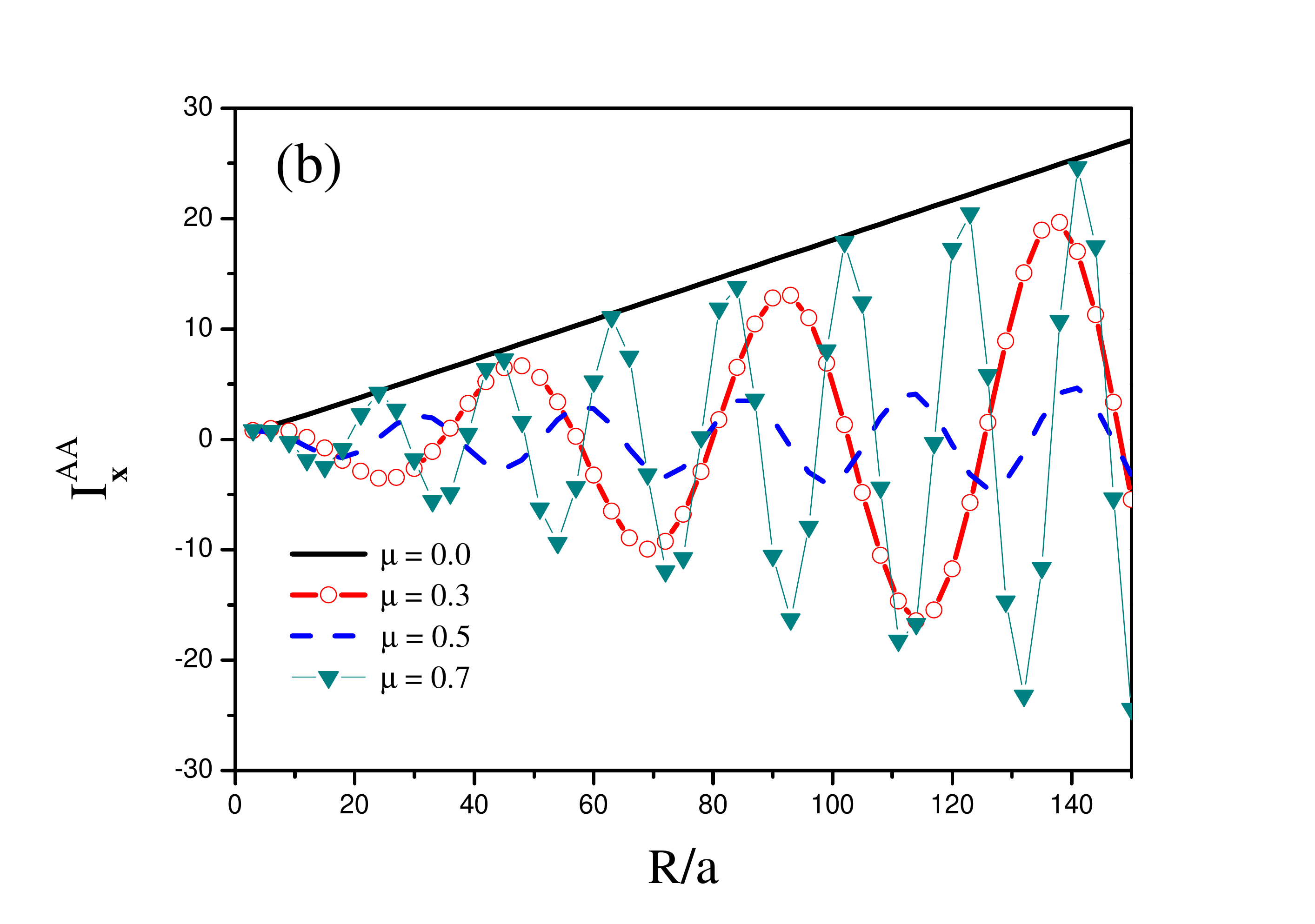}
\caption{(Color online) (a) The integral $I_x$ versus distance between two impurities,
when both impurities are on the same sublattice for doped graphene with the chemical potential
$0.5$ eV and various value of $\mu_p$ in units of eV. (b) Same as (a) but for fixed $\mu_p=0.5$ eV
and various value of the chemical potential.
\label{fig:AAIxEF1} }
\end{figure}

The behavior of the perpendicular components $I_x$ for $\mu\neq 0$
is also different from their linear behavior of the undoped case,
as it is shown in Fig.~\ref{fig:AAIxEF1}(a) for the fixed value of
$\mu=0.5$ eV and different values of $\mu_p$. In this case
$I_x(R)$ exhibits oscillations with a linearly increasing
amplitude whose slope is proportional to $|\mu-\mu_p|$.
Fig.~\ref{fig:AAIxEF1}(b) is the same as Fig.
\ref{fig:AAIxEF1}(a), but this time $\mu_p$ is fixed and $\mu$
changes.

\begin{table*}[t!]
    \begin{center}
\begin{tabular}{l|c|r}
\hline\hline
\vline ~ Chemical potential & Spin polarization & Coupling of strength interaction \vline \\
\hline
\vline ~ $\mu=0$ &  $\mu_p=0$ & $J^{AA}\propto - R^{-3}$ \vline\\
\vline ~ &  & $J^{AB}\propto  R^{-3}$\vline\\
\hline
\vline ~ $\mu\neq 0$ & $\mu_p=0$ & $J^{AA}\propto -\sin(2k_{\rm F}R+\alpha) R^{-2}$\vline\\
\vline ~ & & $J^{AB}\propto \sin(2k_{\rm F}R+\beta) R^{-2}$ \vline\\
\hline
\vline ~ $\mu=0$ & $\mu_p\neq 0$ & $J^{AA}_x\propto -\mu_pR^{-2}$ \vline\\
\vline ~ & & $J^{AA}_z\propto -\sin(2k_{h}R)\mu_pR^{-2}$ \vline\\
\vline ~ & & $J^{AB}_x\propto  \mu_pR^{-2}$ \vline\\
\vline ~ & & $J^{AB}_z\propto + \sin(2k_{h}R)R^{-2} $ \vline\\
\hline
\vline ~ $\mu\neq 0$ & $\mu_p\neq 0$ & $J^{AA}_x\propto -\sin(2k_{\rm F}R)R^{-2} $ \vline\\
\vline ~ & & $J^{AA}_z\propto -(\sin(2x_{\rm F+})+C\sin(2x_{\rm F-})R^{-2}$ \vline\\
\vline ~ & & $J^{AB}_x\propto \sin(2k_{\rm F}R)R^{-2} $ \vline\\
\vline ~ & & $J^{AB}_z\propto (\sin(2x_{\rm F+})+C\sin(2x_{\rm F-})R^{-2} $ \vline\\
\hline\hline
\end{tabular}
\caption{A breakdown of the results on the scaling form of the
RKKY interactions in monolayer graphitic systems. The results for
vanishing spin polarization are reported from
Ref.~[\onlinecite{ref:sherafati}] while the ones for finite spin
polarization denote to present work. It is worthwhile mentioning
that the interaction Hamiltonian is modelled by the Heisenberg
model for $\mu_p=0$ and by the $XXZ$ model for the case that
$\mu_p\neq 0$. We introduce a parameter $k_{\rm h}=\mu_p/v_{\rm
F}$ and $C=(\mu-\mu_p)/(\mu+\mu_p)$ which are given by the
chemical potential as well as the spin polarization. }
\label{table0}
\end{center}
\end{table*}

\section{Summary and Conclusions}\label{sect:conclusion}

In conclusion, we have studied the influence of spin polarization
on RKKY interaction in graphene. With a spin polarization along
the $z$-axis, the induced interaction between two magnetic
impurities is found to be described by an anisotropic XXZ
Hamiltonian with an exchange coupling depending on the distance
$R$ between the impurities and the doping level. For undoped but
spin-polarized graphene, we have found that while the interaction
between the $x$ components of the moments remains constant with
ferromagnetic sign, for the $z$ components it oscillates with the
distance $R$. In unpolarized spin case, the RKKY interaction
induces ferromagnetic correlation between magnetic impurities on
the same sublattice, while anti-ferromagnetic correlation between
the ones on different sublattices~\cite{ref:all,ref:saremi,
ref:sherafati}. The dependence of the interaction on the distance
$R$ between two local magnetic moments, at the Dirac point, is
found to be $R^{-3}$, whereas it behaves as $R^{-2}$ in doped
graphene sheet. Besides $R^{-3}$ dependence of the interaction for
undoped graphene, we show particularly that the interaction
behaves like $R^{-2}$ when the spin polarization is finite.

At finite value of both the chemical potential and the spin
polarization, more complicated behavior of the RKKY coupling can
be occurred. We have found that both components of the interaction
oscillate with $R$. We have explored that for the chemical
potentials $\mu$ close to the polarization $\mu_p$, oscillations
of the RKKY interaction exhibit a beating pattern when the
impurities are located along zigzag or armchair directions. The
two characteristic periods of the beating oscillations are
determined by inverse of the difference and the sum of the
chemical potential and the spin polarization. Since several works
on RKKY interaction in 2D graphene systems are available, a proper
comparison with those results seems to be in order (see Table I).

\section{Acknowledgments}
We are grateful to Jahanfar Abouie for useful discussions. This
work is partially supported by IPM grant.

\end{document}